\documentclass[seceq]{ptptex}
\usepackage{bm}
\usepackage{amsmath}
\usepackage{amssymb}

\usepackage{graphicx}

\title{
Study of exotic hadrons in $s$-wave chiral dynamics%
}

\author{
Tetsuo \textsc{Hyodo}$^{1,}$\footnote{ e-mail address:
hyodo@yukawa.kyoto-u.ac.jp},
Daisuke \textsc{Jido}$^{1}$
and
Atsushi \textsc{Hosaka}$^{2}$
}

\inst{
$^1$Yukawa Institute for Theoretical Physics, 
Kyoto University, Kyoto 606--8502, Japan \\
$^{2}$Research Center for Nuclear Physics (RCNP),
Ibaraki, 567-0047 Japan
}

\abst{
    We study the exotic hadrons in $s$-wave scattering of the 
    Nambu-Goldstone boson with a target hadron based on chiral dynamics. 
    Utilizing the low energy theorem of chiral symmetry, we show that the 
    $s$-wave interaction is not strong enough to generate bound states in 
    exotic channels in flavor SU(3) symmetric limit, although the interaction
    is responsible for generating some nonexotic hadron resonances 
    dynamically. We discuss the renormalization condition adopted in this 
    analysis.
}

\begin{document}

\maketitle

\section{Introduction}

One of the nontrivial issues in hadron physics is almost complete absence of 
flavor exotic hadrons. Experimentally, we have been observing more than 
hundred of hadrons\cite{Yao:2006px}, whose flavor quantum numbers can be 
expressed by minimal valence quark contents of $\bar{q}q$ or $qqq$. The only
one exception is the exotic baryon $\Theta^{+}$ with 
$S=+1$\cite{Nakano:2003qx}, which is composed of at lease five valence 
quarks. In this way, the exotic hadrons are indeed ``exotic'' as an 
experimental fact. On the other hand, there is no clear theoretical 
explanation for the nonobservation of the exotic hadrons. Our current 
knowledge does not forbid to construct four or five quark states in QCD and 
in effective models. Moreover, the multiquark components in nonexotic hadrons
are evident, as seen in the antiquark distribution (or pion cloud) in nucleon
and successful descriptions of some excited hadrons as resonances in 
two-hadron scatterings. In view of these facts, it is fair to say that the 
nonobservation of the exotic hadrons is not fully understood theoretically.

\section{Exotic hadrons in $s$-wave chiral dynamics}

In chiral coupled-channel dynamics, some hadron resonances have been 
successfully described in $s$-wave scattering of a hadron and the 
Nambu-Goldstone (NG) boson\cite{Kaiser:1995eg,Oset:1998it,Oller:2000fj,
Lutz:2001yb}, along the same line with the old studies with phenomenological 
vector meson exchange interaction\cite{Dalitz:1960du,PR155.1649}. It was 
found that the generated resonances turned into bound states in flavor SU(3) 
symmetric limit\cite{Jido:2003cb}. We therefore conjecture that the bound 
states in the SU(3) limit are the origin of a certain class of physical 
resonances, and we examine the possible existence of exotic hadrons as 
hadron-NG boson bound states\cite{Hyodo:2006yk,Hyodo:2006kg}.

The low energy interaction of the NG boson~(Ad) with a target hadron~($T$) in
$s$-wave is given by
\begin{equation}
    V_{\alpha }
    =-\frac{ \omega}{2f^2}C_{\alpha,T}  , 
    \label{eq:WTint}
\end{equation}
with the decay constant $(f)$ and the energy $(\omega)$ of the NG boson. The 
factor $C_{\alpha,T}$ is determined by specifying the flavor representations 
of the target $T$ and the scattering system $\alpha\in T\otimes $Ad:
\begin{equation}
    C_{\alpha,T}
    = - \left\langle 2{\bm F}_{T} \cdot {\bm F}_{\rm Ad}  
     \right\rangle_{\alpha} 
    =  C_2(T)-C_2(\alpha)+3 ,
    \label{eq:WTintfinal}
\end{equation}
where $C_2(R)$ is the quadratic Casimir of SU(3) for the representation $R$.
Eq.~\eqref{eq:WTint} is the model-independent consequence of chiral symmetry,
known as the Weinberg-Tomozawa theorem\cite{Weinberg:1966kf,Tomozawa:1966jm}.

We have written down the general expression of the coupling 
strengths~\eqref{eq:WTintfinal} for arbitrary representations of target 
hadrons in SU(3). In order to specify the exotic channels, we introduced the 
exoticness quantum number, as the number of valence antiquarks to construct 
the given flavor multiplet for the states with positive baryon number. Then 
we find that the Weinberg-Tomozawa interaction in the exotic channels is 
repulsive in most cases, and that possible strength of the attractive 
interaction is given by a universal value
\begin{equation}
    C_{\text{exotic}}=1  , \label{eq:Exoticattraction}
\end{equation}
with $\alpha=[p-1,2]$ for $T=[p,0]$ and $p\geq 3B$\cite{Hyodo:2006yk,
Hyodo:2006kg}.

Next we construct the scattering amplitude with unitarity condition using the
N/D method\cite{Oller:2000fj}. The unitarized amplitude is given by
\begin{equation}
    t_{\alpha}(\sqrt{s})=
    \frac{1}{1-V_{\alpha}(\sqrt{s})G(\sqrt{s})}V_{\alpha}(\sqrt{s}) ,
    \nonumber
\end{equation}
as a function of the center-of-mass energy $\sqrt{s}$. The loop function 
$G(\sqrt{s})$ is regularized by the once subtraction as
\begin{equation}
    G(\sqrt{s})
    =-\tilde{a}(s_0)
    -\frac{1}{2\pi}
    \int_{s^{+}}^{\infty}ds^{\prime}
    \left(
    \frac{\rho(s^{\prime})}{s^{\prime}-s}
    -\frac{\rho(s^{\prime})}{s^{\prime}-s_0}
    \right) ,
    \label{eq:loop}
\end{equation}  
where the phase space integrand is $\rho(s)=2M_{T}\sqrt{(s-s^+)(s-s^-)}/
(8\pi s)$, $s^{\pm}=(m\pm M_{T})^2$, and $m$ and $M_T$ are the masses of the 
target hadron and the NG boson.

In order to determine the subtraction constant $\tilde{a}(s_0)$ and the 
subtraction point $s_0$, we adopt the renormalization condition given in 
Refs.~\citen{Igi:1998gn,Lutz:2001yb},
\begin{equation}
    G(\mu)=0, \quad \mu =M_{T} ,
    \label{eq:regucond} 
\end{equation}
which is equivalent to $t_{\alpha}(\mu)=V_{\alpha}(\mu)$ at this scale. We 
will discuss the implication of this prescription in 
section~\ref{sec:renormalization}. With the condition~\eqref{eq:regucond},
we show that the bound state can be obtained if the coupling 
strength~\eqref{eq:WTintfinal} is larger than the critical value
\begin{align}
    C_{\text{crit}}= \frac{2f^2 }{m\bigl[-G(M_{T}+m)\bigr]} .
    \nonumber
\end{align}
Varying the parameters $m$, $M_T$, and $f$ in the physically allowed region,
we show that the attraction in the exotic 
channels~\eqref{eq:Exoticattraction} is always smaller than the critical 
value $C_{\text{crit}}$. Thus, it is not possible to generate bound states in
exotic channels in the SU(3) symmetric limit.

We would like to emphasize that this conclusion is model independent in the 
SU(3) limit, as far as we respect chiral symmetry. In this study, we only 
consider the exotic hadrons composed of the NG boson and a hadron, so the 
existence of exotic states generated by quark dynamics or rotational 
excitations of chiral solitons is not excluded. In practice, one should bear
in mind that the SU(3) breaking effect and higher order terms in the chiral 
Lagrangian would play a substantial role. Nevertheless, the study of exotic 
hadrons in a simple extension of a successful model of hadron resonances as 
we have done in the present work can partly explain difficulty of observation
of the exotic hadrons.  

\section{Interpretation of the renormalization 
condition}\label{sec:renormalization}

Here we discuss the renormalization condition~\eqref{eq:regucond} in this 
analysis. The interaction kernel $V_{\alpha}(\sqrt{s})$ is constructed from 
chiral perturbation theory so as to satisfy the low energy theorem. The low 
energy theorem also constrains the behavior of the full unitarized amplitude 
$t_{\alpha}(\sqrt{s})$ at a scale $\sqrt{s}=\mu_m$ where the chiral expansion
is valid. Therefore we can match the unitarized amplitude 
$t_{\alpha}(\sqrt{s})$ with the tree level one $V_{\alpha}(\sqrt{s})$ at the 
scale $\mu_m$:
\begin{equation}
    t_{\alpha}(\mu_m)= V_{\alpha}(\mu_m) 
    +V_{\alpha}(\mu_m)G(\mu_m)V_{\alpha}(\mu_m) 
    +\dots
    =  V_{\alpha}(\mu_m) 
    , \label{eq:regucond2}
\end{equation}
This condition determines the subtraction constant such that the loop 
function $G(\mu_m)$ vanishes. This is only possible within the 
region\cite{Hyodo:2006sw}
\begin{equation}
    M_T-m \leq \mu_m \leq M_T+m ,
    \label{eq:region}
\end{equation}
since the loop function has an imaginary part outside this region and the 
subtraction constant is a real number. We consider that by employing this 
renormalization condition, a natural unitarization of the kernel interaction 
based on chiral symmetry is realized. Interestingly, if we apply this 
prescription for the case of the octet baryon target, the subtraction 
constant turns out to be ``natural size'' which was found in the comparison 
with three-momentum cutoff\cite{Oller:2000fj}, and the experimental 
observables in the $S=-1$ meson-baryon channel are successfully reproduced. 

We take $\mu_m =M_T$ in the present study. The dependence on $\mu_m$ within 
the region~\eqref{eq:region} is found that the binding energy of the bound 
state increases if we shift the matching scale $\mu_m$ to the lower energy 
region. This is discussed also in Refs.~\citen{Hyodo:2002pk,Hyodo:2003qa} by 
varying the subtraction constant. The region $\mu_m\leq M_T$ corresponds to 
the $u$-channel scattering. Thus $\mu_m=M_T$ is the most favorable to 
generate a bound state within the $s$-channel regime.

The unitarized amplitudes in the above prescription do not 
always reproduce experimental data. In such a 
case, the subtraction constants $\tilde{a}(s_0)$ should be adjusted in order 
to satisfy experimental data. The subtraction constants determined in this 
way supplement the role of the higher order chiral Lagrangians, which is 
lacking in the kernel interaction. As shown for the $\rho$ meson effect in 
the meson-meson scattering\cite{Ecker:1989te}, the higher order terms may 
contain the effect of the resonances.
Therefore, if the natural condition~\eqref{eq:regucond2} is badly violated, 
one may speculate that the seeds of resonances in the higher order 
Lagrangian, which are possibly the genuine quark states, appear in the 
unitarized amplitude, as in the study of Ref.~\citen{Oller:1998hw}.

In summary, we have argued the following issues. 
\begin{itemize}
    
    \item In order for the unitarized amplitude $t_{\alpha}(\sqrt{s})$ to 
    satisfy the low energy theorem, the loop function should vanish in the 
    region~\eqref{eq:region}.

    \item This requirement can be regarded as a natural unitarization,
    without introducing the effect of resonances in the higher order 
    Lagrangian.

    \item Lower matching scale $\mu_m$ is more favorable to 
    generate a bound state.
    
\end{itemize}
Turning to the problem of exotic hadrons, what we have shown is the 
nonexistence of the exotic bound states with the most favorable condition to 
generate a bound state, without introducing the seed of genuine quark state.

\section*{Acknowledgements}
T.~H. thanks the Japan Society for the Promotion of Science (JSPS) for 
financial support.  This work is supported in part by the Grant for 
Scientific Research (No.\ 17959600, No.\ 18042001, and No.\ 16540252) and by 
Grant-in-Aid for the 21st Century COE "Center for Diversity and Universality in Physics" 
from the Ministry of Education, Culture, Sports, Science and Technology 
(MEXT) of Japan.


\end{document}